\documentclass[twocolumn,prb,showpacs]{revtex4}
\usepackage{graphicx}
\usepackage{dcolumn}
\usepackage{bm}
\usepackage{color}

\begin{document}
\title{Low-lying excitations of the three-leg spin tube using the density-matrix 
renormalization-group method}
\author{S.~Nishimoto$^\ast$}
\affiliation{Max-Planck-Institut f\"ur Physik komplexer Systeme, D-01187 Dresden, Germany}
\author{M.~Arikawa}
\affiliation{Institute of Physics, University of Tsukuba 1-1-1 Tennodai, Tsukuba Ibaraki 305-8571, Japan}
\date{\today}
\begin{abstract}
Using the (dynamical) density-matrix renormalization group method, we study the low-energy 
physics of three-leg antiferromagnetic Heisenberg model where the periodic boundary 
conditions are applied in the rung direction. 
We confirm that the spin excitations are always gapped as long as the intra-ring couplings 
form a regular triangle. From precise finite-size-scaling analyses of the spin gap and 
dimerization order parameter, we also find that the spin gap is collapsed by very small 
asymmetric modulation of the intra-ring couplings. Moreover, the dynamical spin structure 
factors on the intra- and inter-leg correlations are calculated. It is demonstrated that 
the low-lying structure of the inter-leg spectra is particularly affected by the asymmetric 
modulation.
\end{abstract}
\maketitle 

\section{Introduction}

For many years spin ladder systems have attracted much attention not only due to 
the existence of a variety of related materials, e.g., 
Sr$_{n-1}$Cu$_{n+1}$O$_{2n}$ (Refs.~\onlinecite{Azuma94,Kojima95}), 
La$_{4+4n}$Cu$_{8+2n}$O$_{14+8n}$ (Ref.~\onlinecite{Hiroi95}), and 
CaV$_2$O$_5$ (Ref.~\onlinecite{Iwase96}), etc., 
but also as intermediates between one-dimensional (1D) and two-dimensional quantum spin 
physics. It has been confirmed both experimentally and theoretically that 
spin-$\frac{1}{2}$ ladders are gapful for an even number of legs and whereas 
gapless for an odd number of legs when the open boundary conditions (OBC) are applied in 
the rung direction (e.g., as a review, see Ref.~\onlinecite{Dagotto96}). On the other hand, 
if the periodic boundary conditions (PBC) are applied in the rung direction (referred as 
a spin tube) for odd-leg ladders, the spin states are drastically changed by associating 
with the occurrence of frustration; it is known that the system is spontaneously dimerized 
to remove the frustration and all the spin excitations are gapped.~\cite{Schulz96,Kawano97}

At present, there are two experimental candidates for odd-leg spin tubes. 
One of them is vanadium oxide Na$_2$V$_3$O$_7$, which may be regarded as a nine-leg 
Heisenberg spin tube system.~\cite{Millet99} The $^{23}$Na NMR response, dc- 
and ac-magnetic susceptibilities, and the specific heat measurements~\cite{Gavilano05} 
reveal that above 100~K the system is considered as paramagnetic; below 100~K 
most of the localized V magnetic moments ($S=\frac{1}{2}$) form a collection 
of spin-singlet dimers with gaps $\Delta \sim 0-350$~K and the remaining small fraction of 
them forms spin-triplet bound states with gaps $\Delta \sim 0-15$~K; and the degeneracy 
of the triplet ground states is lifted by a phase transition at 0.086 K. 
The low-energy model Hamiltonian of Na$_2$V$_3$O$_7$ has been proposed by some 
theoretical groups. We seem to have reached a consensus that the intra-ring 
exchange interactions are antiferromagnetic. However, inter-ring ones are 
still controversial: the {\it ab initio} microscopic analysis~\cite{Dasgupta05} 
argued that they are frustrated antiferromagnetic and the magnitude is 
much smaller than the intra-ring ones; in contrast, the first-principle 
calculations~\cite{Mazurenko06} estimated them to be ferromagnetic and of 
the same order of magnitude with the intra-ring ones. 

The other experimental candidate is three-leg compound 
[(CuCl$_2$tachH)$_3$Cl]Cl$_2$, which is composed of alternating {(CuCl$_2$tachH)$_3$} 
triangles along the crystallographic $c$ axis.~\cite{Seeber04} The effective 
model has been considered to be a linearly coupled triangle spin rings with 
antiferromagnetic intra-ring couplings and two frustrating antiferromagnetic inter-ring 
couplings. The high-field magnetization measurements suggested that all the couplings 
are of the same order of magnitude; in this situation, it was numerically confirmed 
that the effective model has a spin-gapped ground state.~\cite{Schnack04}

In this paper, motivated by such developments in the field, we study the low-lying 
excitations of three-leg antiferromagnetic Heisenberg spin tube. We assume that 
the fundamental low-energy physics of any odd-leg spin tube can be essentially 
epitomized by that of the three-leg spin tube. So far, several theoretical researches 
have been reported for the three-leg spin tube system: primarily, the bosonization study 
proposed that the three-leg spin tube has a spin-gapped ground state.~\cite{Schulz96} 
It was numerically confirmed and found that the system is completely dimer-ordered 
with a broken translational symmetry.~\cite{Kawano97} 
It has been also suggested that the spin gap is suppressed very rapidly with a lattice 
modulation in the rung direction.~\cite{Sakai05} Additionally, the system 
in a magnetic field~\cite{Cabra97,Cabra98,Citro00,Sato07} and with some kinds of 
frustrated interactions~\cite{Luscher04,Okunishi05,Fouet06} have been examined 
(the latter model is applicable to [(CuCl$_2$tachH)$_3$Cl]Cl$_2$). Thus, 
the ground-state properties of the three-leg spin tube are relatively well understood. 
However, only few theoretical studies have been devoted to the dynamics~\cite{Cabra98,Wang01} 
due to the difficulty of calculating the dynamical quantities. We therefore calculate 
the dynamical spin structure factor using the dynamical density-matrix renormalization 
group (DDMRG) method,~\cite{Jeckelmann02} which has been successfully applied to the 
1D Heisenberg model of late.~\cite{Nishimoto07} Before the dynamical calculation, 
the spin gap and the dimerization order parameter are investigated to provide 
a deeper insight into the ground state. It allows us to precisely analyze the dynamical 
spin structure factor. Based on the results of the static and dynamical quantities, 
we primarily discuss the effect of the lattice modulation in the rung direction on 
the ground-state and low-lying excited-states properties.

This paper is organized as follows. In Sec. II, we define the three-leg 
antiferromagnetic Heisenberg tube and explain the applied methods for 
the calculations. The effective Hamiltonian for the asymmetric case is also 
derived. In Sec. III, we present the calculated results. The 
coupling-strength dependence of the spin gap and the influence of the 
coupling modulation on the low-energy physics are discussed. 
Section V contains summary and conclusions.

\section{Model and Method}

\begin{figure}[t]
    \includegraphics[width= 6.5cm,clip]{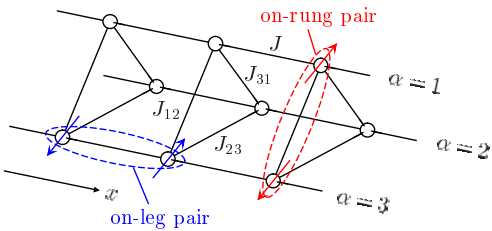}
  \caption{
Lattice structure of three-leg Heisenberg model. Examples of on-leg and on-rung 
spin-singlet pairs are shown.
  }
    \label{lattice}
\end{figure}

\subsection{Hamiltonian}

We consider the three-leg antiferromagnetic Heisenberg model, the Hamiltonian of which 
is given by
\begin{eqnarray}
H = J \sum_{\alpha=1}^3 \sum_i \vec{S}_{\alpha,i} \cdot \vec{S}_{\alpha,i+1} 
+ \sum_{\alpha (\neq \alpha^\prime)} \sum_i J_{\alpha \alpha^\prime} \vec{S}_{\alpha,i} 
\cdot \vec{S}_{\alpha^\prime,i},
\label{hamiltonian}
\end{eqnarray}
where $\vec{S}_{\alpha,i}$ is a spin-$\frac{1}{2}$ operator at rung $i$ and leg $\alpha(=1,2,3)$. 
$J$ ($>0$) is the exchange interaction in the leg direction and $J_{\alpha \alpha^\prime}$ ($>0$) 
is the exchange interaction between the legs $\alpha$ and $\alpha^\prime$ [see Fig.\ref{lattice}]. 
When $J_{\alpha \alpha^\prime}=$ const. ($\forall \alpha, \alpha^\prime$), we call it 
a ``symmetric case'' and set as $J_{12}=J_{23}=J_{31} \equiv J_\perp$; otherwise, 
a ``asymmetric case''. We take $J=1$ as the unit of energy hereafter.

\subsection{DMRG method}

We employ the DMRG technique which is a powerful numerical method for various (quasi) 
1D quantum systems.~\cite{White92} For the calculation of static properties, 
we use the standard DMRG method and the OBC are applied in the leg direction. It enables us to 
calculate ground-state and low-lying excited-states energies as well as static quantities 
quite accurately for very large finite-size systems (up to $\sim {\cal O}(1000)$ sites). 
We are thus allowed to carry out an accurate finite-size-scaling 
analysis for obtaining the energies and quantities in the thermodynamic limit. 
For each calculation, we keep $m = 400$ to $2400$ density-matrix eigenstates in the 
renormalization procedure and extrapolate the calculated quantities to the limit $m \to \infty$. 
We note that the $m$-extrapolation is mandatory in the present system (\ref{hamiltonian}) 
because our DMRG trial state approaches slowly to the exact one with increasing $m$ 
due to very strong spin frustration. In this way, the maximum truncation error, i.e., 
the discarded weight, is less than $1 \times10^{-7}$, while the maximum error in 
the ground-state and low-lying excited states energies is less than $10^{-7}-10^{-6}$. 
For all calculations of the static quantities, we study the ladders with several kinds 
of length $L = 24$ to $312$ and then perform the finite-size-scaling analysis based on 
the system-size dependence of the quantities.

For the calculation of dynamical properties, we use the DDMRG method which is an extension 
of the standard DMRG method and has been developed for calculating dynamical correlation 
functions at zero temperature in quantum lattice models.~\cite{Jeckelmann02} We now apply 
the PBC for both the leg ($x$) and rung ($y$) directions. With the PBC, the system 
size must be restricted practically up to about a hundred but the result is numerically exact 
because the spin operators $\hat{S}^z_{\vec{q}}$ can be precisely defined by
\begin{equation}
\hat{S}^z_{\vec{q}} = \frac{1}{\sqrt{3L}} \sum_l e^{i{\vec{q}}\cdot{\vec{r}}} \hat{S}^z_{\vec{r}},
\label{operator_pbc}
\end{equation}
with momentum ${\vec{q}}=(2\pi z_1/L,2\pi z_2/3)$ for integers $-L/2 < z_1 \le L/2$ and 
$z_2=-1,0,1$.
The sum runs over all sites of the system. Since the exact 
definition of the momentum-dependent operators with the OBC is quite difficult, it would be 
better to choose the PBC for a quantitative estimation of the spectrum. In the DDMRG 
calculation, a required CPU time increases rapidly with the number of the density-matrix 
eigenstates so that we would like to keep it as few as possible; meanwhile, the (D)DMRG 
approach is based on a variational principle so that we have to prepare a `good trial function' 
of the ground state with the density-matrix eigenstates as much as possible. Therefore, 
we keep $m=1200$ to obtain true ground state in the first ten DDMRG sweeps and keep 
$m=400$ to calculate the spectrum for ladders with length $L=24$. In this way, 
the maximum truncation error, i.e., the discarded weight, is about $1 \times10^{-4}$, 
while the maximum error in the ground-state and low-lying excited states energies is 
about $10^{-2}$. 

\subsection{Effective model}

For the symmetric case, in the strong coupling limit $J_\perp \gg J$ the zero-th order 
approximation is obtained by an isolated three-spin triangle. It can be easily diagonalized 
as the higher-energy eigenstates,
\begin{eqnarray}
& & | \uparrow \uparrow \uparrow \rangle, \ \  | \downarrow \downarrow \downarrow \rangle, \nonumber \\
& & \frac{1}{\sqrt{3}}( | \uparrow \uparrow \downarrow \rangle +  
| \uparrow \downarrow \uparrow \rangle +
 | \downarrow \uparrow \uparrow \rangle ), \nonumber \\
 & & \frac{1}{\sqrt{3}}( | \downarrow \downarrow \uparrow \rangle +  
| \downarrow \uparrow \downarrow \rangle +
 | \uparrow \downarrow \downarrow \rangle ),
 \label{higherenergy}
 \end{eqnarray}
with energy $3J_\perp /4$ and the lower-energy eigenstates,
 \begin{eqnarray}
|\uparrow  L \rangle & \equiv & \frac{1}{\sqrt{3}}( | \uparrow \uparrow \downarrow \rangle +  
\omega | \uparrow \downarrow \uparrow \rangle +
\omega^2  | \downarrow \uparrow \uparrow \rangle ), \nonumber \\
|\downarrow  L \rangle & \equiv & \frac{1}{\sqrt{3}}( | \downarrow \downarrow \uparrow \rangle +  
\omega | \downarrow \uparrow \downarrow \rangle +
\omega^2  | \uparrow \downarrow \downarrow \rangle ), \nonumber \\
|\uparrow  R \rangle & \equiv & \frac{1}{\sqrt{3}}( | \uparrow \uparrow \downarrow \rangle +  
\omega^2 | \uparrow \downarrow \uparrow \rangle +
\omega  | \downarrow \uparrow \uparrow \rangle ), \nonumber \\
|\downarrow  R \rangle & \equiv & \frac{1}{\sqrt{3}}( | \downarrow \downarrow \uparrow \rangle +  
\omega^2 | \downarrow \uparrow \downarrow \rangle +
\omega  | \uparrow \downarrow \downarrow \rangle ), 
\label{lowerenergy}
 \end{eqnarray}
with energy $-3J_\perp/4$ where $\omega = \exp(2 \pi i / 3)$. 
From the above four degenerate lower-energy states, we can construct 
the effective Hamiltonian,\cite{Schulz96,Kawano97}
\begin{eqnarray}
H_{\rm eff}^{\rm sym} & = & \frac{J}{3} \sum_{j=1}^L \vec{S}_{j} \cdot  \vec{S}_{j+1} 
\left[1+4( \tau^+_i  \tau^-_{i+1} +  \tau^-_i  \tau^+_{i+1}  ) \right], 
\label{eq.eff1}
\end{eqnarray}
where the chirality operators $\tau^\pm$ are defined as $\tau^+ = | L \rangle \langle R |$ and $\tau^- = | R \rangle \langle L |$, respectively. A previous DMRG study confirmed that the effective Hamiltonian (\ref{eq.eff1}) has a spin gap, which is estimated as $0.277 J$.~\cite{Kawano97}

When the asymmetry is introduced ($J_{12}=J_{31} \neq J_{23}$), the higher-energy states 
(\ref{higherenergy}) are still degenerate and the eigenstates of the anisotropic three-spin 
triangle spin system with energy $(2J_{12}+J_{23})/4$. While, the four degenerate lower-energy 
states (\ref{lowerenergy}) are resolved into the two doublets: ones are
\begin{eqnarray}
| 1 \rangle & = & \frac{| \uparrow L \rangle - \omega | \uparrow R \rangle}{\sqrt{2}}, \nonumber \\
| 2 \rangle & = & \frac{| \downarrow L \rangle - \omega | \downarrow R \rangle}{\sqrt{2}}, 
\label{doublet1}
\end{eqnarray}   
with the eigenenergy $- 3J_{23}/4$, and the others are
\begin{eqnarray}
| 3 \rangle & = & \frac{| \uparrow L \rangle + \omega | \uparrow R \rangle}{\sqrt{2}}, \nonumber \\
| 4 \rangle & = & \frac{| \downarrow L \rangle + \omega | \downarrow R \rangle}{\sqrt{2}}, 
\label{doublet2}
\end{eqnarray}   
with the eigenenergy $-J_{12}+ J_{23}/4$. Thus, the effective Hamiltonian (\ref{eq.eff1}) 
is modified as 
\begin{eqnarray}
H_{\rm eff}^{\rm asym} & = & H_{\rm eff}^{\rm sym} -(J_{12}-J_{23}) \sum_{j=1}^L 
\left[ \omega^2 \tau^+_j  + \omega \tau^-_j \right].  
\label{eq.eff2}
\end{eqnarray}
Since the effect of the asymmetry is written as the crystal-field-type term, 
we suppose that the the physical situation is symmetrical with respect to 
$J_{23}=J_{12}$ as a function of $J_{23}$. Note that the splitting of the four 
lower-energy states (\ref{lowerenergy}) into the doublets (\ref{doublet1}) 
and (\ref{doublet2}) is not complete for nonzero $J$ (see Appendix). In the limit 
that the crystal-field-type term is large, the effective Hamiltonian can be reduced 
to the $S=1/2$ Heisenberg Hamiltonian $J \sum_{j=1}^L \vec{S}_j \cdot \vec{S}_{j+1}$, 
which has no spin gap. 

\section{Results}

\subsection{Spin gap}\label{spingap}

\subsubsection{symmetric case}

\begin{figure}[t]
    \includegraphics[width= 7.0cm,clip]{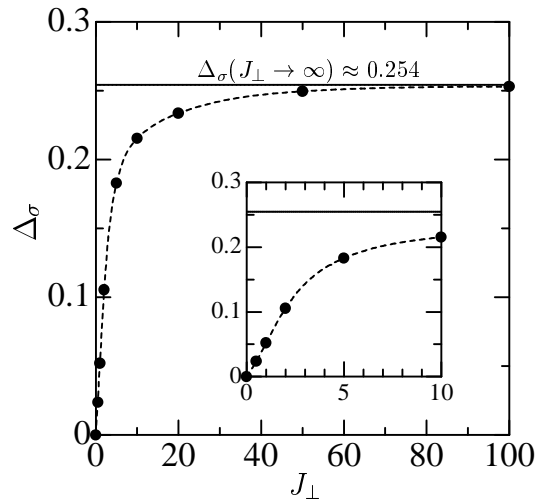}
  \caption{
Spin gap $\Delta_{\rm s}$ as a function of $J_\perp$ in the symmetric case. Inset is 
an extended figure for $J_\perp \le 10$.
  }
    \label{spingap_Jdep}
\end{figure}

First, we study the spin gap in the symmetric case. It is known that 
for all positive $J_\perp$ the system is spontaneously dimerized and all the spin 
excitations are gapped.~\cite{Schulz96,Kawano97} It means that all spins contribute 
to the formation of spin-singlet pairs. If we could pick up a dimerized pair of 
triangles including six spins, three spin-singlet pairs are found: one of them is formed 
in either one of the three legs and the other four spins form a couple of pairs 
in the two rungs. As a result, there exist two types of spin-singlet pairs, namely, 
on-leg and on-rung pairs. Examples of the on-leg and the on-rung spin-singlet pairs 
are shown in Fig.~\ref{lattice}.

It would be very intriguing to see the quantitative dependence of the spin gap on $J_\perp$. 
The spin gap is evaluated by an energy difference between the first triplet excited 
state and the singlet ground state,
\begin{equation}
\Delta_\sigma(L)=E_1(L)-E_0(L), \ \ \ \Delta_\sigma=\lim_{L \to \infty}\Delta_\sigma(L),
\end{equation}
where $E_n(L)$ is the $n$-th eingenenergy ($n=0$ corresponds to the ground state) of 
the system with length $L$, i.e., $L \times 3$ ladder. Note that the number of system 
length must be taken as $L=2l$, with $l (>1)$ being an integer to maintain the total spin 
of the ground state as $S=0$. In Fig.~\ref{spingap_Jdep}, we show the DMRG results of 
the spin gap as a function of $J_\perp$. The plotted values are extrapolated to the 
thermodynamic limit $L \to \infty$ (for example, the extrapolation scheme for $J_\perp=1$ 
is demonstrated in Fig.~\ref{spingap_extrap}). We obtain $\Delta_\sigma=0.254$ in the limit 
of $J_\perp=\infty$. It is rather smaller than a value estimated in Ref.~\onlinecite{Kawano97} 
because the finite-size-scaling analysis is different.

Roughly speaking, the spin gap increases proportionally to $J_\perp$ in the small 
$J_\perp$ ($\lesssim3$) regime and keeps almost constant in the large $J_\perp$ ($\gtrsim10$) 
regime. This behavior can be interpreted in terms of different origin of the spin gap 
for each the $J_\perp$ regime, although the mechanism of gap opening is invariant 
for the entire $J_\perp$ regime. In other words, the spin gap is approximately 
scaled by a binding energy of most weakly bounded spin-singlet pair in the system; 
and, it switches between the on-leg and on-rung pairs at some value of 
$J_\perp (\approx 5)$. A more concrete description is givn in the following paragraph.

For $J \ll J_\perp$, we can easily imagine that the on-rung spin-singlet pairs must 
be bounded more solidly than the on-leg ones. The spin gap is therefore scaled by 
the binding energy of an on-leg pair, i.e., $\Delta_\sigma \propto J$. Accordingly, 
$\Delta_\sigma$ is independent of $J_\perp$ and it is consistent with the constant 
behavior of $\Delta_\sigma$ with $J_\perp$ at $J_\perp \gtrsim 10$. 
On the other hand, the situation is somewhat different for $J_\perp < {\cal O}(J)$: 
the bound state of the on-leg pairs is expected to be more solid than that of 
the on-rung ones. It is because that the system is strongly dimerized with infinitesimally 
small $J_\perp$. The dimerization strength develops abruptly at $J_\perp=0^+$ and 
increases rather slowly with increasing $J_\perp$.~\cite{Nishimoto08} Thus, the spin gap 
is essentially scaled by the binding energy of an on-rung pair. In addition, we may assume 
that the binding energy of the on-rung pair is proportional to $J_\perp$ in the small 
$J_\perp$ regime, by analogy with that of the two-leg Heisenberg system.~\cite{Tsvelik95} 
Now therefore, the spin gap is scaled by $J_\perp$, i.e., $\Delta_\sigma \propto J_\perp$, 
which is consistent to a linear behavior of $\Delta_\sigma$ with $J_\perp$ at 
$J_\perp \lesssim 3$. Note that the derivative $\partial \Delta_\sigma/\partial J_\perp$ 
is very small ($\sim 0.053$) due to strong spin frustration among the intra-ring spins. 
Consequently, a crossover between the constant $\Delta_\sigma$ region and 
the proportional $\Delta_\sigma$ region is seated not at $J_\perp \approx 1$ but 
around $J_\perp \approx 5$. The existence of this crossover can be also confirmed with 
studying the $J_\perp$ dependence of the dynamical spin structure factor. It will 
be discussed in Sec.~\ref{SES}.

\subsubsection{asymmetric case}

\begin{figure}[t]
    \includegraphics[width= 7.0cm,clip]{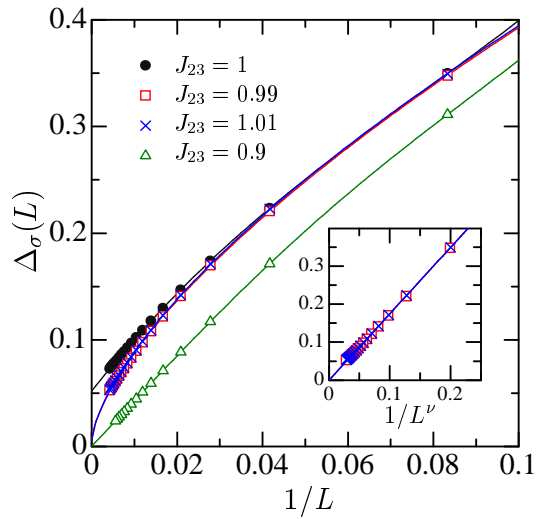}
  \caption{
DMRG results of the spin gap as a function of $1/L$ for several values of $J_{23}$ with 
fixed $J_{12}=J_{31}=1$. Inset: rescaled spin gap as a function of $1/L^\nu$ with 
$\nu=0.649$ and $0.647$ for $J_{23}=0.99$ and $1.01$, respectively.
  }
    \label{spingap_extrap}
\end{figure}

Of particular interest is the evolution of the spin gap onto an asymmetric modulation of 
the intra-ring couplings. Previously, it has been suggested that the spin gap is suppressed 
rapidly by imposing the asymmetry and there exists the Berezinskii-Kosterlitz-Thouless 
(BKT) type transition between gapped and gapless phases at a finite asymmetric 
modulation.~\cite{Sakai05} 
In order to take the asymmetric modulation into account, we vary the value 
of $J_{23}$ from unity with keeping $J_{12}=J_{31}=1$ in our model (\ref{hamiltonian}). 
The system-size dependence of the spin gap for several values of $J_{23}$ is shown 
in Fig.~\ref{spingap_extrap}. For the symmetric case, $J_{23}=1$, $\Delta_\sigma(L)$ can 
be readily extrapolated to $1/L \to 0$ and $\Delta_\sigma=0.052(1)$ is obtained. 
For a relatively large asymmetric case, e.g., $J_{23}=0.9$, we can see that the spin gap is 
obviously extrapolated to zero in the thermodynamic limit.

Let us then consider a small asymmetry by making $J_{23}$ only $1 \%$ smaller from the 
symmetric case, i.e., we set $J_{23}=0.99$. As seen in Fig.~\ref{spingap_extrap}, 
$\Delta_\sigma(L)$ behaves quite similarly to that of the symmetric case for small 
systems ($L \lesssim 50$); however, the deviation comes into the open around 
$L = 50$ and increases rapidly as $1/L$ decreases. Even at a rough estimate, 
$\Delta_\sigma(L)$ seems to be extrapolated 
to a much smaller value at $1/L \to 0$ than that in the symmetric case. For more precise 
extrapolation, a good fitting function ought to be chosen. As shown in the inset of 
Fig.~\ref{spingap_extrap}, we find that $\Delta_\sigma(L)$ can be scaled better by 
$L^{-\nu}$ ($\nu<1$) than by $L^{-1}$ in the case of small asymmetry. This rescaling 
allows us a reasonably performance of the finite-size-scaling analysis to the thermodynamic limit. 
The spin gap is thus obtained as zero within the margin of error $\sim \pm 1 \times 10^{-3}$. 
We also find that the data for $J_{23}=1.01$ are quantitatively the same as those 
for $J_{23}=0.99$, as expected from Eq.~(\ref{eq.eff2}). These results could suggest that 
the spin gap vanishes as soon as a rather (or infinitesimally) small asymmetry in the 
intra-ring exchange couplings is introduced. This is contrary to the BKT-type transition 
suggested in Ref.~\onlinecite{Sakai05}. For further support of our statement, 
we evaluate a dimerization order parameter, which quantifies the presence or 
absence of long-range dimer order indicating the spin-gapped ground state, 
in the following subsection.

We should comment on the finite-size-scaling analysis. The fitting function 
$L^{-\nu}$ ($\nu<1$) may seem to be unusual: however, it is reasonable if we 
take into consideration the fact that the finite-size system will be always dimerized 
due to the Friedel oscillation. As a consequence of the dimerization, the spin gap 
indeed ``opens'' in any finite-size system even if the spin gap closes in the 
thermodynamic limit. We need to remove this ``anomalous'' finite-size effect 
caused by the Friedel oscillation to obtain a correct value in the thermodynamic limit. 
Assuming the dimerization strength due to the Friedel oscillation decays as 
$L^{-\beta}$ ($\beta>0$),~\cite{White02} the spin gap would be scaled as 
$L^{-2\beta/3}$ on the analogy of the results for the 1D spin-Peierls Heisenberg 
model.~\cite{Cross79} Therefore, we can justify the finite-size-scaling analysis 
with the fitting function $\Delta_\sigma(L) \propto L^{-\nu}$ ($\nu=2\beta/3$). 
Incidentally, this finite-size-scaling analysis works only for gapless cases; actually, 
the data for the symmetric case cannot be fitted with $\Delta_\sigma(L) \propto L^{-\nu}$.

\subsection{\label{DOP}Dimerization order parameter}

\begin{figure}[t]
    \includegraphics[width= 6.5cm,clip]{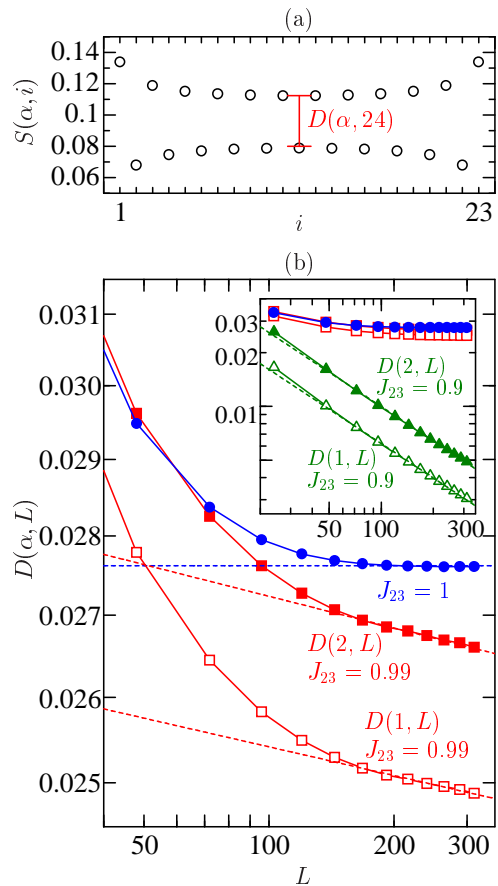}
  \caption{
(a) Friedel oscillation in the nearest-neighbor spin-spin correlations of the system with length 
$L=24$ for a symmetric case $J_\perp=1$. (b) Log-log plots of the amplitude $D(\alpha,L)$ as 
a function of the system length $L$ for $J_{23}=1, 0.99,$ and $0.9$ with fixed $J_{12}=J_{31}=1$.
The dotted lines are the fitting functions $D(\alpha,L) \propto L^{-\gamma}$ ($\gamma \ge 0$).
}
    \label{singletOP}
\end{figure}

Next, we evaluate the dimerization order parameter which indicates the presence or absence 
of long-ranged dimerized state. When the spin gap opens, the system has to be dimerized 
along the leg direction, i.e, the ground state has to be a spin-Peierls one. Therefore, 
a disappearance of this order parameter in the thermodynamic limit corresponds to a collapse 
of the spin gap. In order to know whether the spin gap disappears with small asymmetry, 
we study the dimerization order parameter in the vicinity of the symmetric case.

Because the translational symmetry is broken due to the Friedel oscillation 
under the application of the OBC, the dimerized state is directly observable with 
the DMRG method. We are interested in the formation of alternating 
spin-singlet pairs in the leg direction, so that we calculate the nearest-neighbor 
spin-spin correlations,
\begin{equation}
S(\alpha,i) = -\left\langle \vec{S}_{\alpha,i} \cdot \vec{S}_{\alpha,i+1} \right\rangle,
\end{equation}
where $\left< \cdots \right>$ denotes the ground-state expectation value. 
In Fig.~\ref{singletOP}(a), we show an example of the Friedel oscillation appearing in the 
nearest-neighbor spin-spin correlations of the system with length $L=24$ for a symmetric case 
$J_\perp=1$. (In the symmetric case, the results for all values of $\alpha$ are the same.) 
Generally, the Friedel oscillations in the center of the system decay as a function of 
the system length. If the amplitude at the center of the system
\begin{equation}
D(\alpha,L) = \left|S(\alpha,L/2) - S(\alpha,L/2+1)\right|
\end{equation}
persists for arbitrarily long system length, it corresponds to a long-range dimerization order 
which indicates the spin-Peierls ground state. We thus define the dimerization order 
parameter as
\begin{equation}
D(\alpha)=\lim_{L \to \infty} D(\alpha,L).
\end{equation}
In Fig.~\ref{singletOP}(b), we show the log-log plots of the amplitude $D(\alpha,L)$ as 
a function of the system length $L$ for several values of $J_{23}$ with fixed 
$J_{12}=J_{31}=1$. For the symmetric case, $J_{23}=1$, the derivative 
$-\partial \log D(\alpha,L)/ \partial \log L$ appears to diminish gradually with increasing 
$L$ and $D(\alpha,L)$ saturates at a value, i.e., $D(\alpha)=0.0276(2)$ in the large $L$ limit. 
It signifies the long-range dimerization order, which is consistent with the existence of 
finite spin gap. For reference, we obtain $D(\alpha)=0.0147(5)$ in the large $L$ limit 
at $J_\perp=0.5$ and possibly this value may be proportional to $J_\perp$ in the small 
$J_\perp$ regime.

Let us then study how the order parameter is affected by the asymmetric modulation. 
For a relatively large asymmetric case, $J_{23}=0.9$, the log-log plot of $D(\alpha,L)$ with $L$ 
represents a straight line, i.e., $D(\alpha,L) \propto L^{-0.64}$ ($\forall \alpha$), over 
all ranges of $L$ or at least $L \gtrsim 50$ [see the inset of Fig.~\ref{singletOP}(b)]. 
We thus find a power-law decay of the dimerization order parameter as a function 
of $L$. Since a power-low decay with the distance indicates the absence of long-range 
order, the disappearance of the spin gap is confirmed. We also see that the amplitude 
$D(1,L)$ is smaller than the others $D(2,L)[=D(3,L)]$ though their decay lengths are 
the same. It is so because the reduction of $J_{23}$ prevents the spin-singlet pairs from 
forming on the rungs between the leg $\alpha=2$ and $\alpha=3$; accordingly, the formation of 
the pairs on the leg $\alpha=1$ is strongly suppressed. Note that the (nearly) perfect straight 
line fit of $\log D(\alpha,L)$ versus $\log L$ could be concerned with the (almost) linearly 
scaled spin gap with $1/L$. 

\begin{figure}[t]
    \includegraphics[width= 8.5cm,clip]{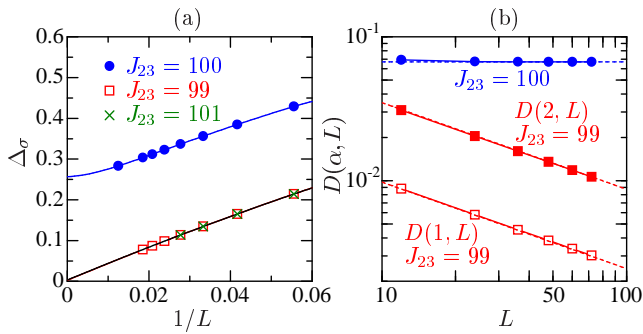}
  \caption{
(a) Spin gap $\Delta_\sigma(L)$ as a function of $1/L$ for several values of $J_{23}$ 
with $J_{12} = J_{31} = 100$. The solid lines are the polynomial fit. (b) Log-log plots of 
the amplitude $D(\alpha,L)$ as a function of $L$. The dotted lines are the fitting 
functions $D(\alpha,L) \propto L^{-\gamma}$. $\gamma$ is estimated as $0.601$ for 
$J_{23}=99$.
}
    \label{J100}
\end{figure}

We now turn to the case with a small asymmetric modulation, $J_{23}=0.99$. 
The derivative of $-\log D(\alpha,L)$ with $\log L$ decreases with increasing $L$, 
as seen in the symmetric case. However, it seems to get at a finite value around 
$L \sim 150$ and the plots for $L \gtrsim 150$ can be fitted by a straight line. 
It again implies a power-law decay of the order parameter in the large distance. 
The fitting function is estimated as $D(\alpha,L) \propto L^{-0.022}$ for all $\alpha$, 
of which the slope is much gentler than that in the case of $J_{23}=0.9$. Although 
the decay of the order parameter is very slow, the formation of the long-range 
dimerization order has to be broken down. 
This slow decay may be the reason why the 
finite-size-scaling analysis of the spin gap is quite difficult (see Fig.~\ref{spingap_extrap}). 
Moreover, it must be a good guess that the decay ratio of the dimerization order 
parameter develops continuously from zero at the symmetric case. Therefore, we argue 
that the spin gap vanishes immediately an infinitesimally small asymmetry is introduced.

The collapse of the spin gap with small asymmetry can be seen more evidently in 
the strong coupling regime $J_\perp \gg J$. As an example, we show (a) the spin gap 
as a function of $1/L$ and (b) the dimerization order parameter as a function of $L$ 
for $J_{12}=J_{31}=100$ in Fig.~\ref{J100}. For the symmetric case, $J_{23}=100$, 
the system has a spin gap 
which is estimated as $\Delta_\sigma=0.253$ and the dimerization order parameter 
converges to $D(\alpha)=0.067$ in the $L \to \infty$ limit. Let us now modulate the 
triangle rings only by $1 \%$, namely, $J_{23}=99$ (and $101$). We obviously find that 
the spin gap is extrapolated to zero in the thermodynamic limit and the dimerization 
order parameter decays as a power law with the system size.

\subsection{A simple intuition}

It would be important to provide an intuitive understanding of the collapse of the 
spin gap with a small asymmetry. Let us consider an isolated triangle Heisenberg ring. 
We assume the three coupling constants to be $K$, $K^\prime$, and $K$, corresponding 
to $J_{12}$, $J_{23}$, and $J_{31}$ in our model, respectively. For $K=K^\prime$, 
the ground state of the ring is four-fold degenerate and the spins are completely 
frustrated. When $K \neq K^\prime$, the degenerate states are splitted into two 
states and the spin gap $\Delta$ opens. If $K^\prime>K$ ($K>K^\prime$), 
a spin-singlet (spin-triplet) pair with $\Delta=K^\prime-K$ ($\Delta=K-K^\prime$) 
between two sites coupled by $K^\prime$. The spin-singlet state corresponds to 
the doublet ones $| 1 \rangle$, $| 2 \rangle$ [Eq,(\ref{doublet1})]; whereas, 
the triplet state corresponds to the doublet ones $| 3 \rangle$, $| 4 \rangle$ 
[Eq,(\ref{doublet2})]. We thus find that a bound state is stabilized for $K \neq K^\prime$. 
We now go back to the three-leg tube. Originally, the spontaneous dimerization is vital 
to form the bound states in the symmetric case. As a results, the spin degrees of 
freedom are quenched and the spin frustration is weakened. However, if the asymmetry 
is introduced, the bound state is naturally formed in each triangle as mentioned above. 
Consequently, the dimer order is no longer necessary. By the analogy with results of 
the single triangle, the binding energy of the bound state in each triangle is scaled 
by $|J_{12}-J_{23}|$ (or $|J_{31}-J_{23}|$) in the asymmetric case. This is consistent 
with the fact that the effective Hamiltonian (\ref{eq.eff2}) includes the effect of 
the asymmetry only as the crystal-field-type term.  

\subsection{\label{SES}Spin excitation spectra}

\subsubsection{symmetric case}

\begin{figure}[t]
    \includegraphics[width= 8.0cm,clip]{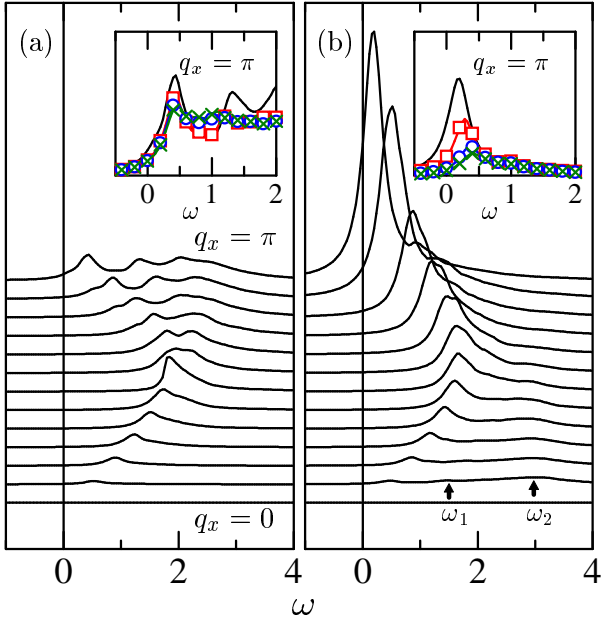}
  \caption{
Dynamical spin structure factor $S(\vec{q},\omega)$ at the symmetric case $J_\perp=1$ 
for (a) $q_y=0$ and (b) $q=\frac{2}{3}\pi$. The system size is fixed at $L=24$ and the broadening 
$\eta=0.1$ is introduced. Insets: $S(\vec{q},\omega)$ at $q_x=\pi$ for $J_\perp=1$ (no symbol), 
$2$ (squares), $5$ (circles), and $10$ (crosses).
  }
    \label{specsym}
\end{figure}

Finally, we study the dynamical spin structure factor to investigate the low-energy 
excitations. For the symmetric case, the spin structure factor is defined as
\begin{eqnarray}
\nonumber
S(\vec{q},\omega) = \sum_n 
\langle \psi_0 \big|\hat{S}^z_{-\vec{q}}\big| \psi_n \rangle
\langle \psi_n \big|\hat{S}^z_{\vec{q}}\big| \psi_0 \rangle \\
\times \delta(\omega-E_n+E_0),
\label{sqsym}
\end{eqnarray}
where $| \psi_n \rangle$ is the $n$-th eigenstate with the eigenenergy $E_n$. 
The operator $\hat{S}^z_{\vec{q}}$ ($q_y=0, \pm \frac{2}{3}\pi$) is the Fourier 
transformation of the spin operator $\hat{S}^z_i$ at site $i$ and given by 
Eq.(\ref{operator_pbc}) when the PBC are applied in the leg direction. 
In $J_\perp=0$ where the system consists of three separated chains, the ground 
state of the whole system is the direct product of the ground state of three 
chains and the dynamical spin structure factors are equivalent to those of 
the 1D Heisenberg model;~\cite{Muller81,Bougourzi96,Karbach97} thus, the spectra 
for $q_y=0$ and $q_y=\frac{2}{3}\pi$ are equivalent. When a finite $J_\perp$ is taken 
into account, the frustration arises among the intra-ring spins. In Fig.~\ref{specsym}, 
we show the DDMRG results of $S(\vec{q},\omega)$ at $J_\perp=1$ for a ladder 
with length $L=24$. A pronounced peak is found at $\vec{q}=(\pi,\frac{2}{3}\pi)$. 
It implies that the spins approximately form a (nearly) $120^\circ$ structure in 
each triangle ring and an antiferromagnetic correlation is dominant along the leg 
direction. On the other hand, the peaks around $\vec{q}=(\pi,0)$ are fairly 
suppressed as compared with the case of $J_\perp=0$. 

For finite $J_\perp$, all the spin excitations are gapped. The position of the 
lowest-lying peak at $\vec{q}=(\pi,0)$ [$\vec{q}=(\pi,\frac{2}{3}\pi)$] essentially 
corresponds to the energy loss to break an on-leg [on-rung] spin-singlet pair.  
Thus, the lowest-excitation energy at either $\vec{q}=(\pi,0)$ or 
$\vec{q}=(\pi,\frac{2}{3}\pi)$ provides the spin gap $\Delta_\sigma$. 
For example, since the on-leg spin-singlet pair has larger binding energy than 
the on-rung one at $J_\perp=1$, the lowest-lying peak of $\vec{q}=(\pi,\frac{2}{3}\pi)$ 
is located at lower frequency than that of $\vec{q}=(\pi,0)$ in Fig.~\ref{specsym}. 
As a result, the position of the largest peak in $\vec{q}=(\pi,\frac{2}{3}\pi)$ 
stands at the lowest spinon excitation whose frequency corresponds to the spin gap. 
The magnitude relation of the binding energies between the on-leg and the on-rung 
pairs switches at $J_\perp \approx 5$ as mentioned in Sec.~\ref{spingap}, 
so that the momentum $\vec{q}$ giving the lowest excitation is also expected to 
change at $J_\perp \approx 5$. 

Let us therefore investigate the evolution of $S(\vec{q},\omega)$ at 
$\vec{q}=(0,\frac{2}{3}\pi)$ and $\vec{q}=(\pi,\frac{2}{3}\pi)$ with $J_\perp$. 
The results are shown in the insets of Fig.~\ref{specsym}. Both the position 
and the weight of the lowest-lying peak at $\vec{q}=(0,\frac{2}{3}\pi)$ are not 
that much sensitive to $J_\perp$, which reflects that the dimerization strength 
is almost saturated above $J_\perp=1$; whereas, at $\vec{q}=(\pi,\frac{2}{3}\pi)$ 
the position of the peak shifts towards higher frequencies and the weight goes 
down with increasing $J_\perp$. We then find that the peak position of 
$\vec{q}=(\pi,\frac{2}{3}\pi)$ comes up with that of $\vec{q}=(0,\frac{2}{3}\pi)$ 
at $J_\perp \approx 5$. It means that the momentum of the lowest-lying peak is 
changed from $\vec{q}=(\pi,\frac{2}{3}\pi)$ to $\vec{q}=(0,\frac{2}{3}\pi)$ at 
$J_\perp \approx 5$. In fact, this change of the momentum corresponds to 
the switch in the origin of the spin gap from the on-rung pair to the on-leg one.

We now refer a couple of the other significant features: one is the appearance of 
two bands around $\omega=1.5$ and $3.0$ in the $q_y=\frac{2}{3}\pi$ spectra, 
which are denoted as $\omega=\omega_1$ and $\omega_2$ in Fig.~\ref{specsym}, respectively. 
Those bands are associated with internal excitation on each triangle ring. Considering 
an isolated triangle ring, there are three states; namely, a spin-$\frac{3}{2}$ quadruplet 
with energy $\frac{3}{4}$ [Eq.~(\ref{higherenergy})] and two degenerate spin-$\frac{1}{2}$ 
doublets with energy $-\frac{3}{4}$ [Eq.~(\ref{lowerenergy})]. Hence, when $J=0$ 
the excitation spectra show two flat bands at $\omega_1=0$ and 
$\omega_2=\frac{3}{2}J_\perp$. Thus, we can easily guess the spin structure factor 
in the strong coupling regime $J_\perp \gg J$: the spectra of the low-energy excitations 
[$\omega \sim {\cal O}(J)$] are essentially the same as those of the 1D spin-Peierls (SP) 
Heisenberg model~\cite{Pytte74,Tsvelik92,Castilla95,Haas95,Watanabe99}; and single peaks 
associated with the intra-ring excitation exist around $\omega_2=J+\frac{3}{2}J_\perp$ 
for all of $q_x$. The other feature is the reduction of the apparent width of 
the continuum in the $q_y=\frac{2}{3}\pi$ spectra, relative to that of 
the 1D Heisenberg model. In other words, the peaks at the lower edge of the spectra 
are rather the $\delta$-function like. It possibly reflects the local excitation of 
the spinon bound state on each dimer, as in the spectra of the 1D SP Heisenberg 
model.~\cite{Watanabe99}

\subsubsection{asymmetric case}

\begin{figure}[t]
    \includegraphics[width= 8.0cm,clip]{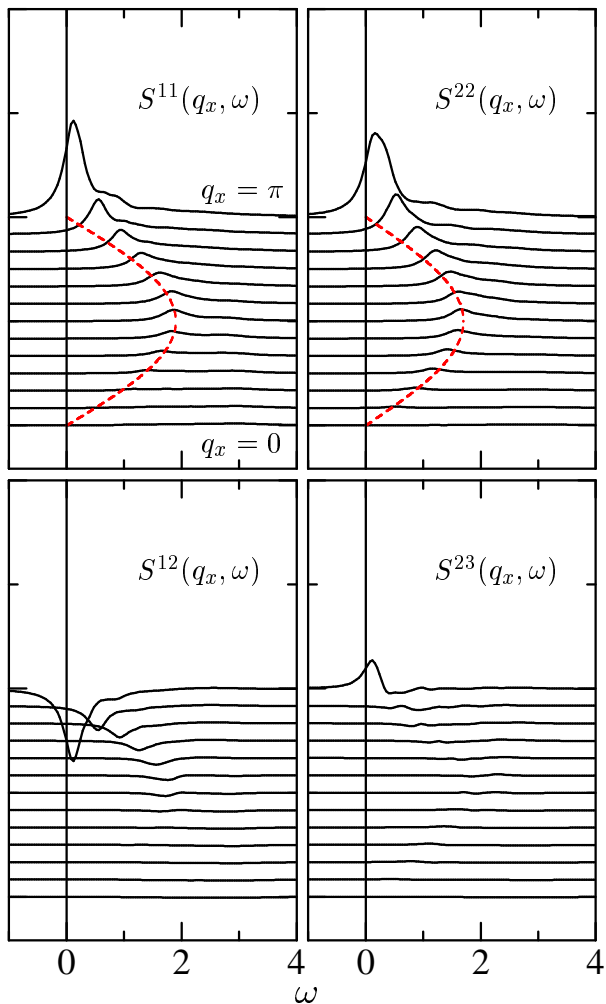}
  \caption{
Dynamical spin structure factor $S^{\alpha\beta}(q_x,\omega)$ at the asymmetric case 
$J_{12}=J_{31}=1$ and $J_{23}=0.9$. The system size is fixed at $L=24$ and the broadening 
$\eta=0.1$ is introduced. The dashed lines denote the fitting of the lowest-lying peaks 
with $\omega \sim \sin q_x$.
  }
    \label{sqasym}
\end{figure}

We then examine the dynamical spin structure factor in the asymmetric case. 
Particularly, we focus on the change of the spectra with the asymmetry. 
Since the translation symmetry in the rung direction is no longer present 
for the asymmetric case, we redefine the dynamical spin structure factor as 
\begin{eqnarray}
\nonumber
S^{\alpha\alpha^\prime}(q_x,\omega) = \sum_\nu 
\langle \psi_0 \big|\hat{S}^z_{\alpha,-q_x}\big| \psi_\nu \rangle
\langle \psi_\nu \big|\hat{S}^z_{\alpha^\prime,q_x}\big| \psi_0 \rangle \\
\times \delta(\omega-E_\nu+E_0)
\label{corr2}
\end{eqnarray}
with
\begin{equation}
\hat{S}^z_{\alpha,q_x} = \frac{1}{\sqrt{L}} \sum_x e^{iq_xx} \hat{S}^z_{\alpha,x}, 
\label{operator2_pbc}
\end{equation}
where there are four independent combinations of $\alpha$ and $\alpha^\prime$, i.e., 
$(\alpha,\alpha^\prime)=(1, 1)$, $(2, 2)$, $(1, 2)$, and $(2, 3)$. 

We here consider the system with a $10 \%$ asymmetric modulation, i.e., $J_{12}=J_{31}=1$ 
and $J_{23}=0.9$. The DDMRG results of $S^{\alpha\beta}(q_x,\omega)$ for a ladder 
with length $L=24$ are shown in Fig.~\ref{sqasym}. The spin gap should close at 
$q_x=0$, $\pi$ and the intra-leg spin structure factors $S^{11}(q_x,\omega)$ and 
$S^{22}(q_x,\omega)$ are basically similar to that of the 1D Heisenberg 
model.~\cite{Muller81,Bougourzi96,Karbach97} The lower edge of the spectra 
is well-fitted with a function $\omega = c \sin q_x$, and $c$ is estimated as $1.9$ 
and $1.7$ for $S^{11}(q_x,\omega)$ and $S^{22}(q_x,\omega)$, respectively. 
One of the most noticeable deviation from the spectra of the 1D Heisenberg model 
is the rapid decrease of the spectral weights $S(q_x)$ with distance from $q_x=\pi$. 
It is because that the spin fluctuations are suppressed by the exchange interaction 
between the legs and the antiferromagnetic correlations in the leg direction are 
`longer-ranged'. Since the leg $\alpha=1$ is more strongly coupled with 
the neighboring legs than the leg $\alpha=2$, the spectral weights of 
$S^{11}(q_x,\omega)$ at small $q_x$ are fewer than those of $S^{22}(q_x,\omega)$. 
For example, the similar feature has been confirmed in the spin 
structure factor of the two-dimensional Heisenberg model which has 
an antiferromagnetically ordered ground state.~\cite{Manousakis91} 

Let us turn to the inter-leg spin structure factors $S^{12}(q_x,\omega)$ and 
$S^{23}(q_x,\omega)$. These factors themselves represent changes of the 
spectral features derived from the asymmetric modulation, because 
they must be zero in the symmetric case. In $S^{12}(q_x,\omega)$, negative-weighted 
peaks appear around $q_x=\pi$ . It means that the the antiferromagnetic 
correlation between the legs $\alpha=1$ and $\alpha=2$ increases from the frustrated 
$120^\circ$ spin structure for the symmetric case. Whereas in $S^{23}(q_x,\omega)$, 
smaller peaks with positive weights appear around $q_x=\pi$. It signifies the emergence 
of the ferromagnetic correlation between the legs $\alpha=2$ and $\alpha=3$, 
which is caused by the superexchange interaction via the leg $\alpha=1$. 
With increasing the asymmetric modulation, the low-energy physical properties 
of the three-leg ladder with the PBC in the rung direction seem to quickly 
approach to those of a three-leg ladder with the OBC.~\cite{Reigrotzki94}

\section{SUMMARY}

We study the low-lying excitations of the three-leg antiferromagnetic Heisenberg tube 
with the (D)DMRG method. For the symmetric case, we argue that the spin gap is 
scaled by the binding energy of the on-rung spin-singlet pair in the weak-coupling regime 
($J_\perp \lesssim 3$); whereas, the on-leg spin-singlet pair in the strong-coupling regime 
($J_\perp \gtrsim 10$). We then take an asymmetric modulation of the intra-ring exchange 
couplings into account. For small asymmetries, precise finite-size-scaling analyses of 
the spin gap and dimerization order parameter are carried out. Based on the results, 
we suggest that the spin gap vanishes as soon as an infinitesimally small asymmetry is 
introduced. 

Furthermore, we calculate the dynamical spin structure factors. 
In the symmetric case, the low-energy spectra are essentially the same as those of 
the 1D SP Heisenberg model. Note, however, that additional peaks associating the 
intra-ring excitation exist. In the asymmetric case, the intra-leg spectra are basically 
similar to those of the 1D Heisenberg model. They are hardly affected by the asymmetric 
modulation except the spin-gap closing at the band edge. On the other hand, 
the inter-leg spectra are profoundly affected even by a small asymmetric modulation. 
An enhancement of ferromagnetic or antiferromagneic correlations between two legs 
can be clearly seen. It means that the low-energy physics of the three-leg Heisenberg 
tube approaches quickly to that of the non-tube three-leg ladder with increasing 
the asymmetry.

\acknowledgments

We thank T.Sakai, K.Okunishi, K.Okamoto, C.Itoi, M.Sato, Y.Otsuka, Y. Hatsugai, S. Tanaya, 
and T. Takimoto for useful discussions. MA is supported by 
the University of Tsukuba Research Initiative.

\appendix
\section*{appendix}

In this Appendix, we consider how effectively the four degenerate states 
(\ref{lowerenergy}) are resolved to either of the doublets (\ref{doublet1}) 
or (\ref{doublet2}) in the ground state. If only configurations containing 
one of the doublets are included in the asymmetric case, the Hilbert space 
to be considered can be much reduced. This reduction is exact only in the limit 
of $J \to 0$ and thus we check the validity of the reduction numerically 
for nonzero $J$.

\begin{figure}[t]
    \includegraphics[width= 5.0cm,clip]{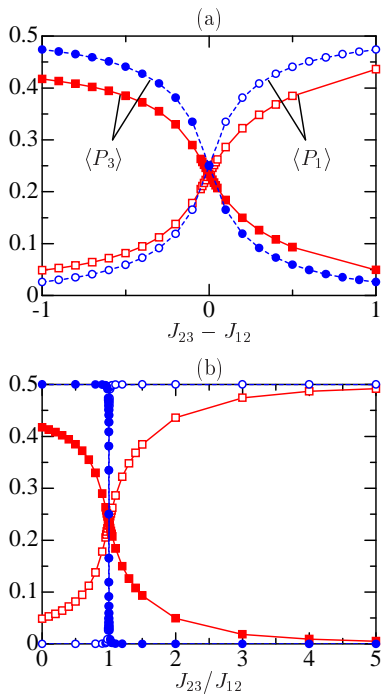}
  \caption{
Ground-state expectation values of the projection operators 
$\langle P_1 \rangle$ ($=\langle P_2 \rangle$) (empty) and 
$\langle P_3 \rangle$ ($=\langle P_4 \rangle$) (filled) for $J_{12}=J_{31}=1$ (squares) 
and $100$ (circles).  
  }
    \label{proj_fig}
\end{figure}

When $|J_{12}-J_{23}| \gg J$, the ground state of each triangle is approximately 
expressed as a linear combination of the states $| 1 \rangle$ and $| 2 \rangle$ 
for $J_{23}>J_{12}$; whereas, $| 3 \rangle$ and $| 4 \rangle$ for $J_{23}<J_{12}$. 
However, the four states are equally likely `mixed' around $J_{12}=J_{23}$ for 
nonzero $J$. We here prepare projection operators $P_i=| i \rangle \langle i |$ ($i=1 \cdots 4$). 
In Fig.~\ref{proj_fig}(a), the ground-state expectation values of the projection operators 
$\langle P_1 \rangle$ ($=\langle P_2 \rangle$) and $\langle P_3 \rangle$ ($=\langle P_4 \rangle$) 
as a function of $|J_{12}-J_{23}|$ are shown for a couple of cases $J_{12}=J_{31}=1$ 
and $100$. We can see that the dominance of the doublet state exchanges continuously 
between $| 1 \rangle$,$| 2 \rangle$ and $| 3 \rangle$,$| 4 \rangle$ at 
$|J_{12}-J_{23}| \lesssim {\cal O}(J/2)$.
It is confirmed for $J_{12}=J_{31}=100$ that either $\langle P_1 \rangle$ or 
$\langle P_3 \rangle$ is $\sim 1/2$ and the other is $\sim 0$ at $|J_{12}-J_{23}| \gtrsim J/2$, 
where the reduction to the doublet of the Hilbert space  in each triangle must be applicable. 
Since we have the relation $-P_1 -P_2 +P_3 + P_4 = \omega^2 \tau^+ +  \omega \tau^- $,
the values $\langle P_1 \rangle$ and $\langle P_3 \rangle$ are symmetric on the reflection 
against the line $J_{23}-J_{12}=0$, which is consistent with the crystal-field-type term 
in the effective Hamiltonian (\ref{eq.eff2}). 
We also plot the same results as a function of $J_{23}/J_{12}$ in Fig.~\ref{proj_fig}(b). 
For $J_{12}=J_{31}=1$, the situation seems to be more complex but the properties 
are expected to be qualitatively similar to those in the strong-coupling limit 
$J_{12},J_{23} \gg J$.

\end{document}